\documentclass[conference]{IEEEtran}
\IEEEoverridecommandlockouts

\usepackage{amsmath,amssymb,amsfonts}
\usepackage{mathtools}
\usepackage[linesnumbered,ruled]{algorithm2e}
\usepackage{algpseudocode} 
\usepackage{subfigure}
\usepackage{enumitem}

\usepackage{threeparttable}
\usepackage{booktabs}
\usepackage{makecell}  
\usepackage{arydshln} 
\usepackage{multirow}
\usepackage{adjustbox}
\usepackage{stackengine}
\newcommand\xrowht[2][0]{\addstackgap[.5\dimexpr#2\relax]{\vphantom{#1}}}

\usepackage[dvipsnames]{xcolor}
\definecolor{background}{HTML}{FFFFF5}
\definecolor{edge}{HTML}{87C2B1}
\definecolor{bisque}{HTML}{B86500}
\definecolor{myblue}{HTML}{2A4597}

\usepackage[colorlinks=true,citecolor=black, linkcolor=black, urlcolor=myblue]{hyperref}
\usepackage{cleveref}

\usepackage[most]{tcolorbox}
\newtcolorbox{mybox}{colback=background!35,%background 40% opacity
colframe=edge,% black frame colour
width=\columnwidth,% total width
arc=2mm,
auto
outer
arc
}

\usepackage[]{collab}

\usepackage{xspace}
\def\tool{{Maat}\xspace}

\newcommand{\red}[1]{\textcolor{red}{#1}}

\newcommand{\bisque}[1]{\textcolor{bisque}{#1}}

\begin{document}
\title{\tool: Performance Metric Anomaly Anticipation for Cloud Services with Conditional Diffusion}

\author{
  \IEEEauthorblockN{
    Cheryl Lee\IEEEauthorrefmark{2},
    Tianyi Yang\IEEEauthorrefmark{2},
    Zhuangbin Chen\IEEEauthorrefmark{1}\IEEEauthorrefmark{3}\thanks{\hspace{-2ex}\IEEEauthorrefmark{3}Zhuangbin Chen is the corresponding author.},
    Yuxin Su\IEEEauthorrefmark{1}, and
    Michael R. Lyu\IEEEauthorrefmark{2}
  }

  \IEEEauthorblockA{\IEEEauthorrefmark{2}Department of Computer Science and Engineering, The Chinese University of Hong Kong, Hong Kong, China.\\
    Email: cheryllee@link.cuhk.edu.hk, \{tyyang, lyu\}@cse.cuhk.edu.hk}

  \IEEEauthorblockA{\IEEEauthorrefmark{1}School of Software Engineering, Sun Yat-sen University, Zhuhai, China.\\
    Email: \{chenzhb36, suyx35\}@mail.sysu.edu.cn}
}

\maketitle

\begin{abstract}
    Ensuring the reliability and user satisfaction of cloud services necessitates prompt anomaly detection followed by diagnosis.
    Existing techniques for anomaly detection focus solely on real-time detection, meaning that anomaly alerts are issued as soon as anomalies occur.
    However, anomalies can propagate and escalate into failures, making faster-than-real-time anomaly detection highly desirable for expediting downstream analysis and intervention.
    This paper proposes Maat, the first work to address anomaly anticipation of performance metrics in cloud services.
    Maat adopts a novel two-stage paradigm for anomaly anticipation, consisting of metric forecasting and anomaly detection on forecasts.
    The metric forecasting stage employs a conditional denoising diffusion model to enable multi-step forecasting in an auto-regressive manner. 
    The detection stage extracts anomaly-indicating features based on domain knowledge and applies isolation forest with incremental learning to detect upcoming anomalies.
    Thus, our method can uncover anomalies that better conform to human expertise.
    Evaluation on three publicly available datasets demonstrates that Maat can anticipate anomalies faster than real-time comparatively or more effectively compared with state-of-the-art real-time anomaly detectors.
    We also present cases highlighting Maat's success in forecasting abnormal metrics and discovering anomalies.

\end{abstract}


\section{Introduction}
With the wave of cloud computing, recent years have witnessed a dramatic expansion in the scale and complexity of cloud services~\cite{LogSurvey, IncidentManagement, QoSSurvey}.
However, cloud systems are prone to performance anomalies due to the complexity of the underlying cloud system and service interaction structure~\cite{CloudChallenge}.
Even worse, minor anomalies could magnify their impacts and escalate into serious failures, worsening end users' satisfaction and resulting in severe revenue implications~\cite{HCloud, EffortAID}.
Thus, guaranteeing service reliability in cloud systems hinges on effectively managing performance anomalies.
As cloud infrastructure scales, monitoring data increases exponentially, making manual anomaly detection unfeasible and costly.

Hence, tremendous efforts have been expended toward the automation of cloud service assurance~\cite{EffortAlert, EffortIncident, OmniAnomaly, adsketch, Hades}. 
These methods rely on run-time data, inclusive of logs, traces, and various performance metrics such as CPU usage, memory consumption, and I/O loads. 
However, existing studies all focus on real-time detection through alert issuance after suspicious monitoring data are detected, followed by downstream tool provision of relevant data on anomaly localization, root cause analysis, and countermeasures~\cite{EffortAlert}. The downstream analysis is time-consuming, during which anomalies can propagate, cascade and evolve into significant failures.
Therefore, there is a pressing need for advanced approaches in handling faster-than-real-time (FTRT) anomaly detection to trigger alarms proactively, which can expedite downstream analysis and provide opportunities for intervention to prevent further propagation and revert cloud services to a stable state.

In light of the limitations of real-time anomaly detection, we propose a novel paradigm, anomaly anticipation, to achieve FTRT anomaly detection.
This two-stage paradigm consists of metric forecasting and anomaly detection based on the forecasts combined with real-time observations.
Yet, achieving this paradigm of anomaly anticipation is not straightforward. We identify three key challenges:
1) \textit{Aggressive forecasts}: Widely-used forecasting models~\cite{TCN, gru, DeepVAR, Attn} tend to make conservative forecasts that are restricted to the range of the previous values, making it less likely to forecast anomalous patterns.
2) \textit{Visible forecasting values}: Besides a near-feature binary output (normal or not), the anomaly anticipator needs to forecast concrete and numerical performance metric values to allow for downstream data-driven analysis. 
3) \textit{Detection of interest}: The detected anomalies of a completely data-driven and black-box detector may not correspond to the interest of cloud services.
The detector should allow the integration of domain expertise to facilitate the understanding of anomalies and gain trust from operational engineers.

By addressing these challenges, we propose \tool, the first performance \underline{M}etric \underline{A}nomaly \underline{A}n\underline{T}icipation approach for cloud services.
\tool employs a conditional denoising diffusion model for the forecasting stage, enabling the autoregressive generation of metrics. It extracts temporal dependencies and cross-metric relations as the condition and samples the next-time metric vector.
Denoising diffusion models, inspired by non-equilibrium thermodynamics, have shown great success in high-quality generations, especially in text, images, audio, etc~\cite{DDPM, DiffScore, DiffYang}.
Motivated by the generative progress of diffusion models, \tool provides a novel application of the diffusion model to address cloud computing challenges by learning probability distributions of historical monitoring metrics to generate a next-time sample of the learned distribution conditioned on learned dependencies. 
In this way, \tool realizes multi-step forecasting in an autoregressive manner and enables accurate forecasts even though the ground truths are abnormal signals with a small probability, thereby addressing the first and second challenges.
The detection stage of \tool establishes a set of anomaly-indicating features and applies incremental learning with isolation forest~\cite{IF}. 
The expertise-dependent feature extraction and the tree-based model facilitate interpretable anomaly detection.
Hence, \tool represents a significant innovation in FTRT anomaly anticipation in cloud services by integrating the forecasting and detection processes, while avoiding over-complexities and total agnosticism.

We evaluate \tool on three widely-used public datasets.
The experimental results demonstrate that \tool can achieve comparatively or more optimal FTRT anomaly anticipation compared with state-of-the-art real-time anomaly detectors.
In particular, \tool improves the F1-score by a significant margin (15.15\%$\sim$79.95\%) on average, and it is worth noting that only \tool can issue anomalies before their occurrence.
Additionally, the real-time detector component of \tool also exhibits superiority over the competitors (19.17\%$\sim$86.76\% higher in an average F1-score). 
Besides, our forecaster outperforms recent advances in probabilistic forecasting of multivariate time series.
Specific cases also illustrate \tool's success in forecasting abnormal metrics and distinguishing anomalies.

In summary, the main contributions of this paper are:
\begin{itemize}[leftmargin=10pt, topsep=0pt]
    \item We are the first to propose and formulate anomaly anticipation as a new paradigm, which can be easily extended to achieve the anticipation of root causes, fault diagnosis, and other reliability challenges in cloud computing.
    \item We propose \tool, the first framework to proactively anticipate cloud-service performance anomalies based on a novel conditional diffusion model. We are the first to apply diffusion models to tackle software engineering challenges. The implementation code has been publicly released on \href{https://github.com/BEbillionaireUSD/Maat}{https://github.com/BEbillionaireUSD/Maat}.
    \item We conduct extensive experiments to demonstrate that \tool, as an FTRT anomaly anticipation approach, is comparatively or more effective than state-of-the-art real-time anomaly detectors, thereby enabling alerts in advance and saving much time for downstream analysis.
\end{itemize}
\section{Background and Problem Formulation}\label{sec:background}
This section introduces performance metrics, denoising diffusion models, and the formulation of anomaly anticipation.

\subsection{Background}
Herein, we introduce the data and technique, i.e., the denoising diffusion model, based on which this paper works.

\subsubsection{Cloud-service Performance Metrics}
\textit{Performance metrics} are uniformly sampled real-valued time series measuring the system status, categorized into three main areas generally: resource utilization (e.g., CPU usage, memory usage), workload performance (e.g.,  error rate, throughput), and service level agreement (e.g., uptime, response time). They are responsive to performance changes and can provide meaningful insights into cloud reliability.
Usually, a system has multiple monitoring metrics reflecting different aspects of the system's performance, and the number of collected values of each metric is its \textit{length}.

\subsubsection{Denoising Diffusion Model}\label{sec:DDPM}
Denoising diffusion models~\cite{DDPM} are novel generative models inspired by non-equilibrium thermodynamics.
They add noise to the input (i.e, forward) and learn to generate by eliminating the noise (i.e., reverse).
%% Adding noise
In the forward process, Gaussian noise is gradually added to the input $x_0\sim q(x_0) \in \mathbb{R}^{m \times l}$ containing $m$ $l$-length metrics with the approximate posterior:
\begin{equation}\label{eq:posterior}
    q(x_1, \cdots, x_T|x_0) 
    \coloneqq 
    \prod_{t=1}^T q(x_t|x_{t-1})
\end{equation}
Each step of the Eq.~\ref{eq:posterior} is:
\begin{equation}
    q(x_t|x_{t-1}) \coloneqq \mathcal{N}(x^n; \sqrt{1-\beta_n} x^{n-1}, \beta_n \boldsymbol{I})
\end{equation}
where $\beta_1, \cdots, \beta_N \in (0,1)$ are learned or pre-defined variance schedules.
As $x_t$ is only related to $x_{t-1}$, the forward process is parameterized as a Markov chain.
The end, $x_T$, is totally corrupted to be random noise: $p(x_T)=\mathcal{N}(x_T; \boldsymbol{0}; \boldsymbol{I})$.

%% Denoising
In the reverse process, we define a probability density function $p_\theta(x_0)$ to approximate $q(x_0)$ since the original distribution of inputs is unknown. 
Starting from $x_T$, we can restore $x_0$ with learned Gaussian transitions by iteratively calculating:
\begin{equation}\label{eq:p-1-step}
    p_\theta(x_{t-1} | x_t) \coloneqq \mathcal{N}(x_{t-1}; \boldsymbol{\mu}_\theta (x_t, t), \boldsymbol{\Sigma}_\theta (x_t, t))
\end{equation}
where parameters $\theta$ are learnable and shared;
$\boldsymbol{\mu}_\theta: \mathbb{R}^m \times \mathbb{N} \rightarrow \mathbb{R}^m$ and $\boldsymbol{\Sigma}_\theta: \mathbb{R}^m \times \mathbb{N} \rightarrow \mathbb{R}^+$ are functions of the corrupted input $x_t$ and the forward step's index $t \in \mathbb{N}_+$.
Thus, the joint distribution is:
\begin{equation}
    p_\theta(x_0,\cdots,x_{T-1}|x_T)\coloneqq p(x_T) \prod_{t=1}^T p_\theta (x_{t-1} | x_t)
\end{equation}

Naturally, the training objective is to minimize the KL-divergence between distributions $p$ and $q$. With Jensen’s inequality and the speeding-up parameterization proposed in~\cite{DDPM}, the objective is derived to be:
\begin{equation}\label{eq:loss}
    \min_\theta \mathcal{L} \coloneqq \min_\theta \mathbb{E}_{x_0 \sim q(x_0), \boldsymbol{\epsilon}\sim \mathcal{N}(\boldsymbol{0}, \boldsymbol{I}),t} \|\boldsymbol{\epsilon}-\boldsymbol{\epsilon_\theta}(x_t,t)\|_2^2
\end{equation}
where $\boldsymbol{\epsilon}$ denotes the noise added to the input in the forward process and $\boldsymbol{\epsilon_\theta}$ is a trainable model taking $x_t = \sqrt{\bar{\alpha}_t}x_0+(1-\bar{\alpha}_t)\boldsymbol{\epsilon}$ and the index $t$ as inputs.

Hence, $\boldsymbol{\mu}_\theta(x_t, t)$ and $\boldsymbol{\Sigma}_\theta(x_t, t)$ in Eq.~\ref{eq:p-1-step} are rewritten by (Detailed proofs are given by~\cite{DDPM}):
\begin{equation}\label{eq:param}
\begin{aligned}
    \boldsymbol{\mu}_\theta (x_t, t) & \coloneqq \frac{1}{\sqrt{\alpha_t}}\left(x_t - \frac{\beta_t}{\sqrt{1-\bar{\alpha}_t} }\boldsymbol{\epsilon_\theta} (x_t,t)\right),\\ 
    \boldsymbol{\Sigma}_\theta (x_t, t) &\coloneqq \sigma_t^2 \boldsymbol{I}
\end{aligned}  
\end{equation}
where $\tilde{\beta} \coloneqq \frac{1-\alpha_{t-1}}{1-\alpha_t}\beta_t$ if $t>1$; $\tilde{\beta} \coloneqq \beta_1$ if $t=1$.

Once trained, we can sample $x_t$ with any $t$ from the corrupted input $x_T\sim \mathcal{N}(\boldsymbol{0}, \boldsymbol{I})$, so by experimentally setting $\boldsymbol{\Sigma}_t^2$ to be $\frac{1-\bar{\alpha}_{t-1}}{1-\bar{\alpha}_t}$, we can estimate $\hat{x}_0$ from white noise:
\begin{equation}\label{eq:sample}
    x_{t-1}=\boldsymbol{\mu}(x_t, t) + \boldsymbol{\Sigma}_t \cdot \varkappa, \,
    \varkappa \left\{
        \begin{aligned}
            &\sim \mathcal{N}(\boldsymbol{0}, \boldsymbol{I}),\, & t>1 \\
            &=\boldsymbol{0},\, & t=1
        \end{aligned}
    \right.
\end{equation}
Thus, given an arbitrary noise vector serving as a pseudo $x_T$, we can generate a new sample by estimating the $\hat{x}_0$, even if the real $x_0$ does not exist in a generation task.

To sum up, denoising diffusion models attempt to learn the probability distribution of the performance metrics, and the learned distribution enables sampling values as forecasts.

\subsection{Problem Formulation}
This section distinguishes between three tasks: real-time anomaly detection, early anomaly detection and our proposed anomaly anticipation, each presented with its corresponding mathematical formulation.

\subsubsection{Real-time Anomaly Detection}
%% 介绍X[1:t]-->y[t]
Real-time anomaly detection is a far-reaching field in cloud computing. Anomaly detectors aim to discover abnormal signals from observed monitoring metrics reflecting various performance aspects, such as CPU usage, memory, network, and so on. Anomalies are observations containing patterns inconsistent with expectations, historical data patterns, or human knowledge. 
Real-time anomaly detection can be formulated as follows.
Given observed data $x_{[1:t]} \in \mathbb{R}^{m \times t}$ at the time of $t$, where $m$ is the number of metrics, a detector should output $\hat{y}_t \in \{0,1\}$ to denote the existence of anomalies (normal: 0, abnormal: 1). 
A real-time anomaly detector is only possible to capture the anomaly after its occurrence.

\subsubsection{Early Anomaly Detection}\label{sec:background:early}
%% X[1:t] --> y[t+s]
When a threat to cloud services is detected, every second counts. Thus, detecting anomalies (including failures, incidents, and outages) in real-time is a bit late in some scenarios~\cite{eWarn, AirAlert, NTAM}. That is where early anomaly detection comes in, which aims to extract the relationship between observations and the future system status. 
Given current monitoring metrics $x_{[1:t]}$, an early anomaly detector will output $y_{t+s} \in \{0,1\}$ to indicate upcoming anomalies, where $s$ is the number of advanced time steps.

\subsubsection{Anomaly Anticipation}
%% X[1:t] --> hat X[t+1:t+s] --> y[t+s]
We propose anomaly anticipation, a brand-new paradigm to identify anomalies before they occur.
Given the observed data $x_{[1:t]}$, the first phase aims to build a model $f_{1}$ to forecast the near-feature data, i.e., $f_{1}(x_{[1:t]})=\hat{x}_{[t+1:t+s]}$, where $s$ is the number of forecasting time steps. 
Then, another model $f_{2}$ detects anomalies based on the concentration of the observed and forecasted values, i.e., $f_{2}([x_{[1:t]}, \hat{x}_{[t+1:t+s]}])=\hat{y}_t$, where $\hat{y}_t \in \{0,1\}$ indicates whether a current (upcoming) anomaly exists (will occur).

Instead of mining the exclusive relationships between current metrics and upcoming anomalies, anomaly anticipation forecasts predictable metrics and pinpoints abnormal patterns $s$-step in advance. 
The first phase is self-supervised, while the second phase can be modeled as unsupervised or semi-supervised.
Hence, we propose a simple but effective paradigm to decompose a challenging and not thoroughly studied problem into two known problems with in-depth research achievements. 
Note that the FTRT anticipation result is usually poorer than the real-time detection result, as forecasting will inevitably introduce imperfections and anomaly anticipation is much more challenging than detection.

\section{Motivation and Challenges}\label{sec:motivation}
This section first introduces the necessity and feasibility of anomaly anticipation, motivated by which we propose a novel paradigm of anomaly anticipation. Afterward, we discuss the challenges of achieving this paradigm, which calls for our solution for effective anomaly anticipation.

\subsection{Motivation}\label{sec:motivation:motivation}
\subsubsection{Issues of Current Practice}

To our best knowledge, most, if not all, current performance metric-based practices are for real-time anomaly detection, aiming to identify anomalies in a target service~\cite{Hades, OmniAnomaly, adsketch, mtad, dount}.
Though abnormality itself may not directly result in huge losses, it can cause serious failures~\cite{GrayFailure}.
Therefore, anomaly detection, which typically comes before more in-depth analysis, like root cause analysis and fault diagnosis, is given great importance.
Despite encouraging progress, existing real-time detectors lack the ability to alert operation engineers before an anomaly occurs and escalates into a failure, leading to potential disasters during such an alert delay.
Given the complexity of fine-grained diagnosis and recovery, there is a need for speeding up the prerequisite step, i.e., anomaly detection, to save valuable time for downstream slower analysis.

Early anomaly detection ($\S$\ref{sec:background:early}) has gradually gained attention in a broad sense, particularly in predicting incidents~\cite{eWarn}, outages~\cite{AirAlert}, and disk failures~\cite{NTAM, CDEF} for cloud systems. 
These efforts aim to extract the relationship between the current data and near-future status, whether normal or not.
However, before an anomaly arises, there are usually few visible abnormal behaviors, meaning that the relationships are very subtle and unfathomable.
Therefore, existing proactive anomaly detectors are mostly supervised~\cite{AirAlert, NTAM, eWarn, CDEF}.
The lack of high-quality labels and distrust from operation teams have been bottlenecks of these approaches.
Even worse, such approaches only provide binary results without providing concrete information on the underlying abnormal data.
This makes it difficult to analyze the causes earlier and even prevent anomalies proactively, rendering early detection less useful.

In this context, there is a growing need for more advanced anomaly detectors that can not only detect anomalies faster than real-time but also provide insights into why the detector predicts an impending anomaly. It is crucial to provide visible metric values of the upcoming anomalies to enable earlier downstream analysis.

\subsubsection{Anticipatability of Anomalies}
After discussing the necessity of anticipating anomalies, we present a twofold feasibility illustration. 
First, due to the inherent temporal correlation within cloud services, as they adopt similar programming models and underlying systems, forecasting their performance metrics is possible. 
Second, as an anomaly is characterized by a significant deviation from the expected behavior, it must exhibit notable distributional differences from normal metrics, enabling detection within cloud services. 
Hence, considering the temporal correlation and distributional characteristics of performance anomalies in cloud services, the anticipation of performance anomalies is indeed a feasible task. 
The detailed illustration will be presented one by one.

Cloud services, regardless of the adopted architecture, such as vertical~\cite{vertical-architecture}, service-oriented~\cite{soa-architecture}, or microservice~\cite{microservice-architecture}, comprise multiple software components that interact with each other and operate under a defined programming logic, i.e., ``do certain reactions to certain user requests or system events''.
For example, to process a user request ``post a new blog'', a cloud service needs to execute a series of tasks: a) compress the user-uploaded images, b) safely store the compressed image in a backend storage service (e.g., Amazon S3), c) store the text contents to a backend database.
Note that a) is a computation-intensive task, b) and c) are network-intensive.
Since metrics reflect the execution status and the execution logic of the underlying cloud system, processing such a request will incur a peak in CPU usage followed by a peak in network traffic.
Hence, the intrinsic temporal relation between a cloud service's actions leads to the temporal correlation reflected in performance metrics, whose two adjacent periods are logically related.
In summary, the similarity in programming models and underlying infrastructure of cloud services contributes to the temporal correlation of their monitoring metrics.

Anomalous metrics exhibit distributional differences from normal metrics.
Previous investigations~\cite{ADSurvey, YahooSurvey} have identified three types of distributional differences in cloud systems: point anomalies, contextual anomalies, and collective anomalies.
Point anomalies indicate global peaks in application latency or resource usage metrics, while contextual anomalies only indicate local anomalous patterns in specific execution environments. 
Collective anomalies occur when each metric point is normal, but the joint occurrence of metric points becomes abnormal, defined by cumulative observations.
For example, compared with the past high throughput, the unexpected fringe of low throughput value is anomalous.
Both contextual and collective anomalies are rooted in the intrinsic logic of cloud services.
By capturing this logic and the temporal relations of performance metrics, it is possible to foresee the occurrence of contextual and collective anomalous metrics. 
Thus, anomaly anticipation can be achieved by detecting abnormality in forecasted performance metrics.

\begin{mybox}
  \small
  \textbf{Insight}: Due to the commonality of programming models and underlying systems in cloud services, the performance metrics exhibit an intrinsic temporal correlation, making their behaviors predictable.
  By applying anomaly detection techniques to forecasted performance metrics, it is possible to anticipate anomalies in cloud services.
\end{mybox}

\subsection{Challenges and Solutions}
We identify three technical challenges in achieving FTRT anomaly anticipation and provide our point-to-point solutions.

\subsubsection{Aggressive Forecasting}

Anticipating anomalies in cloud services requires accurate forecasted metrics, as forecasting errors can affect the accuracy of anomaly detection. Most forecasting models tend to make conservative forecasts, which restrict the forecasted value within the range of previous values. They are less likely to forecast abnormal signals with low probability. 
To address this issue, probabilistic generative models such as diffusion models are preferred over discriminative models. We also use a smooth L1 loss function to encourage more aggressive forecasts.

\subsubsection{Visible Forecasting Values}

Existing early anomaly detectors only produce a binary output indicating whether an upcoming anomaly will occur, but a binary output is far from sufficient for downstream analysis and preventive decisions, which require concrete data.
To address this challenge, we propose a two-stage forecasting-detecting model that produces numerical performance metrics along with a binary anomaly indicator.
This approach empowers downstream data-driven analysis and proactive actions.

\subsubsection{Detection of anomalies of interest}
Enabling human reasoning about the detector decisions is crucial to winning the trust of engineers as well as facilitating diagnosis and proper preventive responses. 
Totally data-driven methods (unsupervised) identify novel patterns as abnormal, but it does not necessarily correspond to cloud applications. For example, resource usage metrics are excepted to experience peaks during holidays for a traveling-related service.
Hence, we apply feature engineering to the concatenation of observed and forecasted metrics.
Leveraging domain knowledge, we manually define performance-related features of six categories that can indicate anomalies. Afterward, we build up an interpretable tree-based model to distinguish anomalies from normal ones.

\section{Methodology}\label{sec:method}
\begin{figure}[htb]
    \vspace{-0.1in} 
    \centering
        {\includegraphics[width=0.95\linewidth]{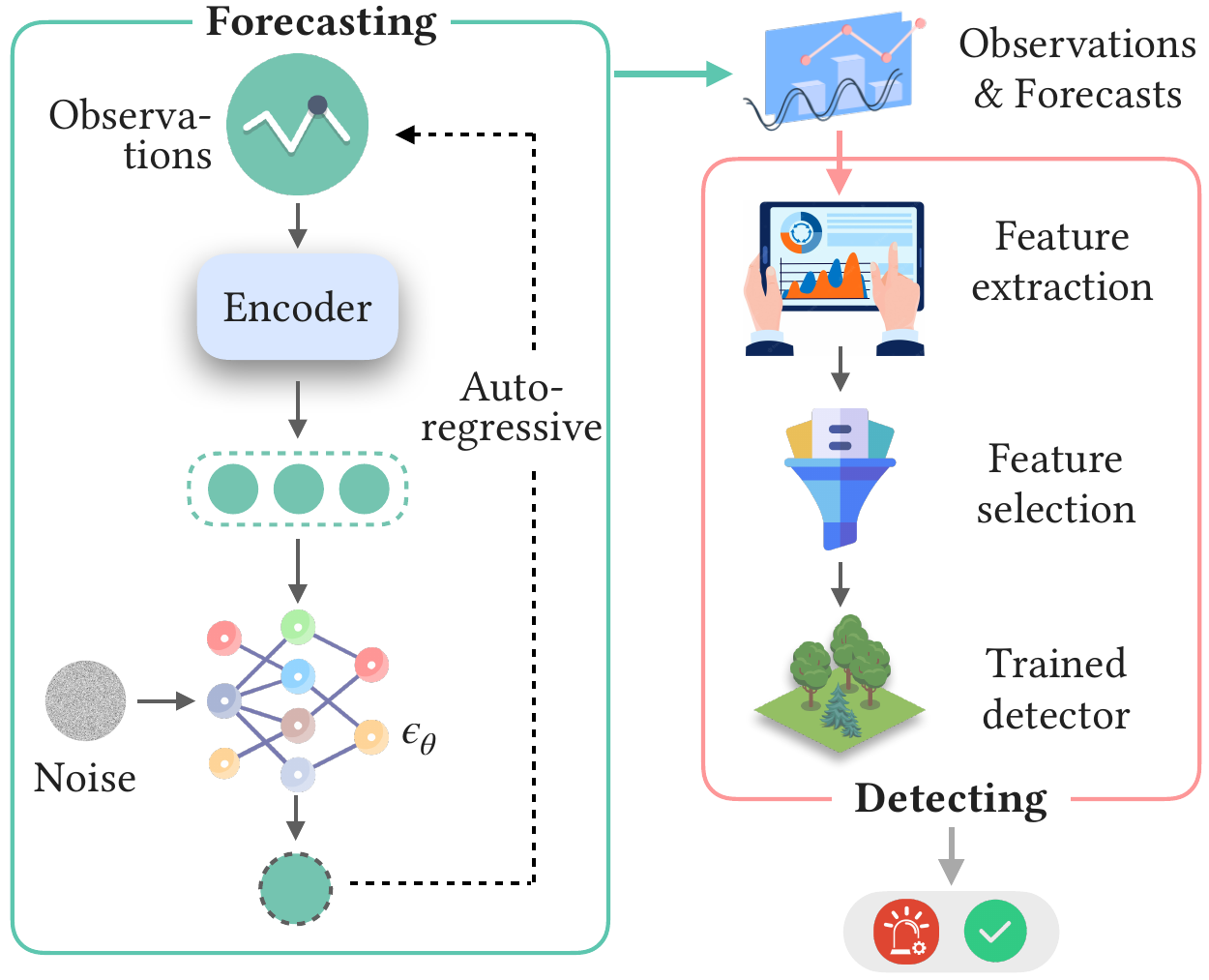}}
    \vspace{-0.1in}
    \caption{Overview of \tool.}
     \vspace{-0.1in} 
    \label{fig:tool}
\end{figure}

Figure~\ref{fig:tool} presents the overview of \tool, an anomaly anticipation framework for cloud services.
It consists of two main components: a conditional diffusion model-based forecaster and a tree-based detector.
The forecaster captures the temporal and cross-metric dependencies among performance metrics by encoding them into high-dimensional embeddings as the condition, based on which a denoising diffusion model forecasts the near-feature performance metrics auto-regressively. 
The detector extracts anomaly-indicating features from the concatenation of observations and forecasts and then employs an incrementally trained isolation forest to anticipate existing or upcoming performance anomalies.

\subsection{Metric Forecasting}
\begin{figure}[htb]
    \centering
        {\includegraphics[width=0.95\linewidth]{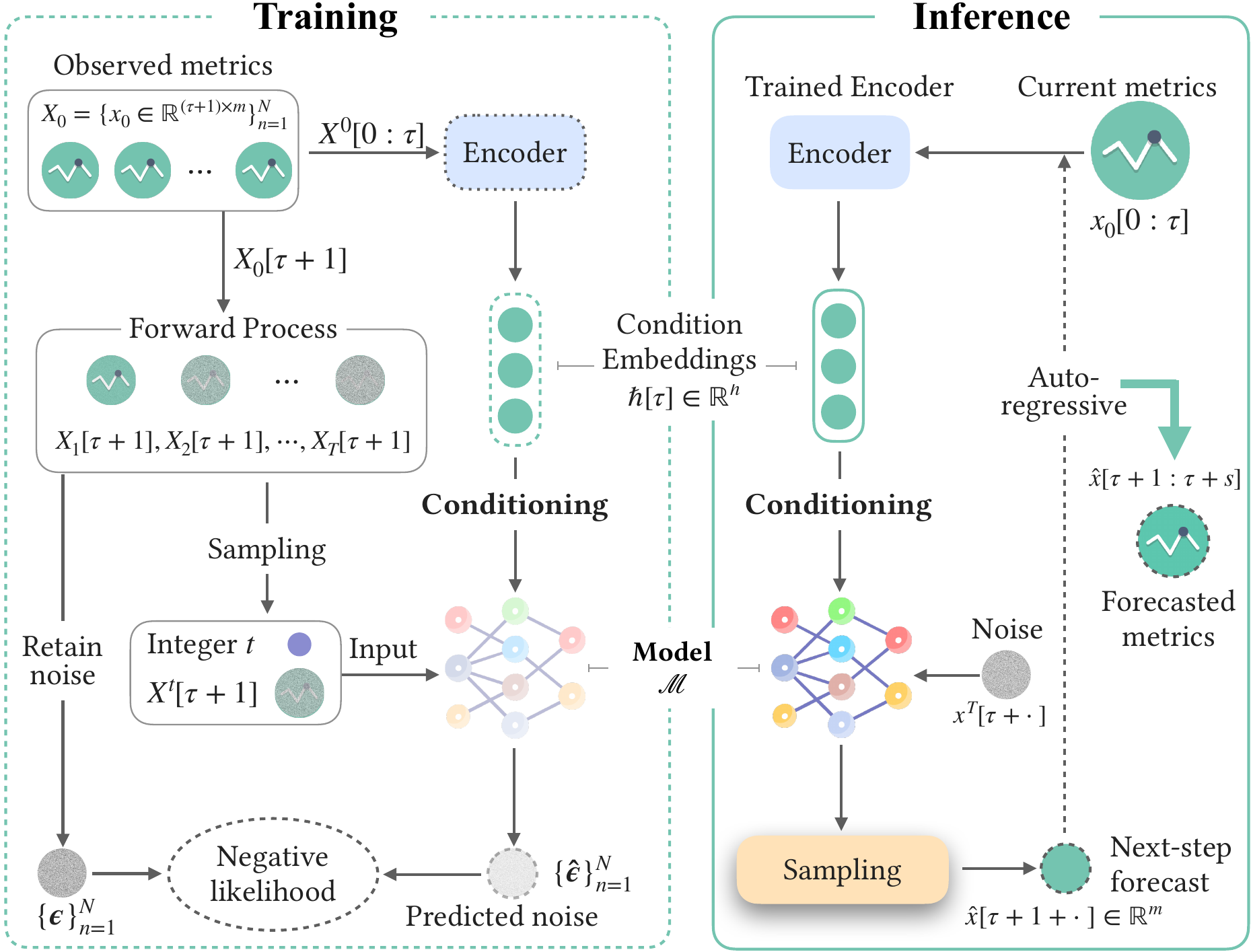}}
    \vspace{-0.1in}
    \caption{Work process of the forecaster based on conditional denoising diffusion model.}
    \label{fig:diffusion}
\end{figure}

This section introduces how we forecast by incorporating conditions to improve upon naive denoising diffusion models, which do not consider the short-term context.
Denoising diffusion models, based on probabilistic generation, are selected over other time series prediction approaches based on discriminative models. They can better handle abrupt changes, volatility shifts, and irregular patterns, so they are suitable for our aggressive forecasting with anomalies.
Moreover, the reasons for incorporating conditions are two-fold.
First, according to the analysis in~\ref{sec:motivation:motivation} and existing studies~\cite{Sage, Seer, Hades}, previous metrics directly influence the following values. 
Second, cross-metric relations also influence performance or resources. For example, a CPU-extensive workload may also eat up much memory.
As plenty of factors may influence the future's metrics, we regard the near-historical observations as conditions, drawing inspiration from successful applications of conditional time series generation~\cite{TimeGrad, MAF, STRIPE}.
Figure~\ref{fig:diffusion} shows the process of metric forecasting. We will introduce conditioning, training, inference, and model $\mathcal{M}$ in detail.

\textbf{Conditioning.} 
We first embed performance metrics into a high-dimensional space by extracting meaningful information via an encoder.
Various neural networks (e.g. temporal convolutional network~\cite{TCN}) or embedding techniques (e.g., Time2Vec~\cite{Time2Vec}) can serve as the encoder to capture temporal dependencies and cross-metric correlations. 
We choose the gated recurrent unit (GRU)~\cite{gru} due to its lightweight architecture and superior performance in signal processing.
Through a multi-layer GRU network with learnable parameters $\theta$, we obtain the metric embeddings, denoted by:
\begin{equation}
    \hbar= \text{GRU}_\theta([x^\tau_0, c^{\tau+1}], \hbar^{\tau-1})
\end{equation}
where $x^\tau_0 \in \mathbf{R}^m$ is the $\tau$-th value of the input $x_0$ with $m$ metrics, and $c^{\tau+1}$ are next-time covariates.

\textbf{Training.} 
Similar to a naive denoising diffusion model aiming to minimize the KL-divergence between $p$ and $q$ (Eq.~\ref{eq:loss}), the training objective conditioned on $\hbar$ is to narrow the gap between $q(z|\hbar)$ and $p_\theta(z|\hbar)$.
In anomaly anticipation, the forecaster should make accurate forecasts even under anomalous contexts, so we encourage aggressive forecasts through smooth convergence by introducing the smooth L1 loss~\cite{FastRCNN} to replace the commonly used L2 loss used in Eq.~\ref{eq:loss}, because smooth L1 loss is more robust and less sensitive to outliers.
Given two vectors $\boldsymbol{x}, \boldsymbol{y} \in \mathbb{R}^N$, smooth L1 loss is
$
    \mathcal{L}_{\text{s}}(\boldsymbol{x}, \boldsymbol{y}) = \frac{1}{N}\sum_{n=1}^N l_n,
$
where 
$$l_n=\left\{
        \begin{aligned}
            &0.5(x_n-y_n)^2, & \text{if}\, |x_n-y_n|<1 \\
            &|x_n-y_n|-0.5,  & \text{otherwise}
        \end{aligned}
    \right.$$
Hence, with $N$ training windows, the training objective in Eq.~\ref{eq:loss} is further reformulated on conditions and simplified as:
\begin{equation}\label{eq:final-loss}
    \min_\theta \mathcal{L} \coloneqq \sum_{\tau=l}^{l+s} \min_\theta \mathbb{E}_{x_0^{\tau}, \boldsymbol{\epsilon},t} \left[\mathcal{L}_{\text{s}}\left(\boldsymbol{\epsilon}-\boldsymbol{\epsilon_\theta}(x_t^{\tau},t, \hbar^{\tau-1}) \right)\right]
\end{equation}
where $l$ is the context length and $s$ is the forecasting length, and the superscript $n$ is omitted for simplicity.

\textbf{Inference.} 
Again, referring to the sampling of a naive diffusion model in Eq.~\ref{eq:sample}, the trained model forecasts the next-step value by sampling from the learned distribution conditioned on $\hbar$:
\begin{equation}\label{eq:forecast}
    x_{t-1}^{\tau+1} = \frac{1}{\sqrt{\alpha_t}}\left(x_t^{\tau+1} - \frac{\beta_t}{\sqrt{1-\bar{\alpha}_t} }\boldsymbol{\epsilon_\theta} (x^{\tau+1}_t,t, \hbar^\tau)\right) + \boldsymbol{\Sigma}_t \cdot \varkappa
\end{equation}
where the initial input $x_T^{\tau+1}\sim \mathcal{N}(\boldsymbol{0}, \boldsymbol{I}) \in \mathbb{R}^m$ is an arbitrary noise vector.
The forecasts are inputted autoregressively to the encoder so that we can obtain multi-step ($s$-step) forecasts $\hat{x}^{[l+1:l+s]}$, given the $l$-length observation $x_0^{[1:l]}$. 

Moreover, as the model is probabilistic, we sample more than once and then adopt the mean for smoothing. We can also apply maximum or minimum for more aggressive forecasts. It is up to the application scenario, e.g., CPU usage is relatively stable, while I/O metrics can fluctuate dramatically.

\textbf{Model} $\boldsymbol{\mathcal{M}.}$ 
Having decided the diffusion model, how to predict the noise vector, $\boldsymbol{\hat{\epsilon}} \leftarrow \boldsymbol{\epsilon_\theta}(\cdot)$, remains.
We address this by viewing the forecaster as a special auto-encoder, with both the input and output being noises. 
Thus, $\boldsymbol{\epsilon_\theta}(\cdot)$ should be a conditional generative decoder.
Particularly, we borrow Conditional PixelCNN from the domain of image generation to serve as $\boldsymbol{\epsilon_\theta}(\cdot)$ because it is a powerful decoder capable of efficiently generating novel high-quality images with pixel-level precision, especially considering that anomaly anticipation requires intricate detail and pattern generation.
Its flexibility in incorporating different types of conditional information, such as latent embeddings provided by other neural networks, also encourages our selection.

\begin{figure}[htb]
    \centering
        {\includegraphics[width=0.9\linewidth]{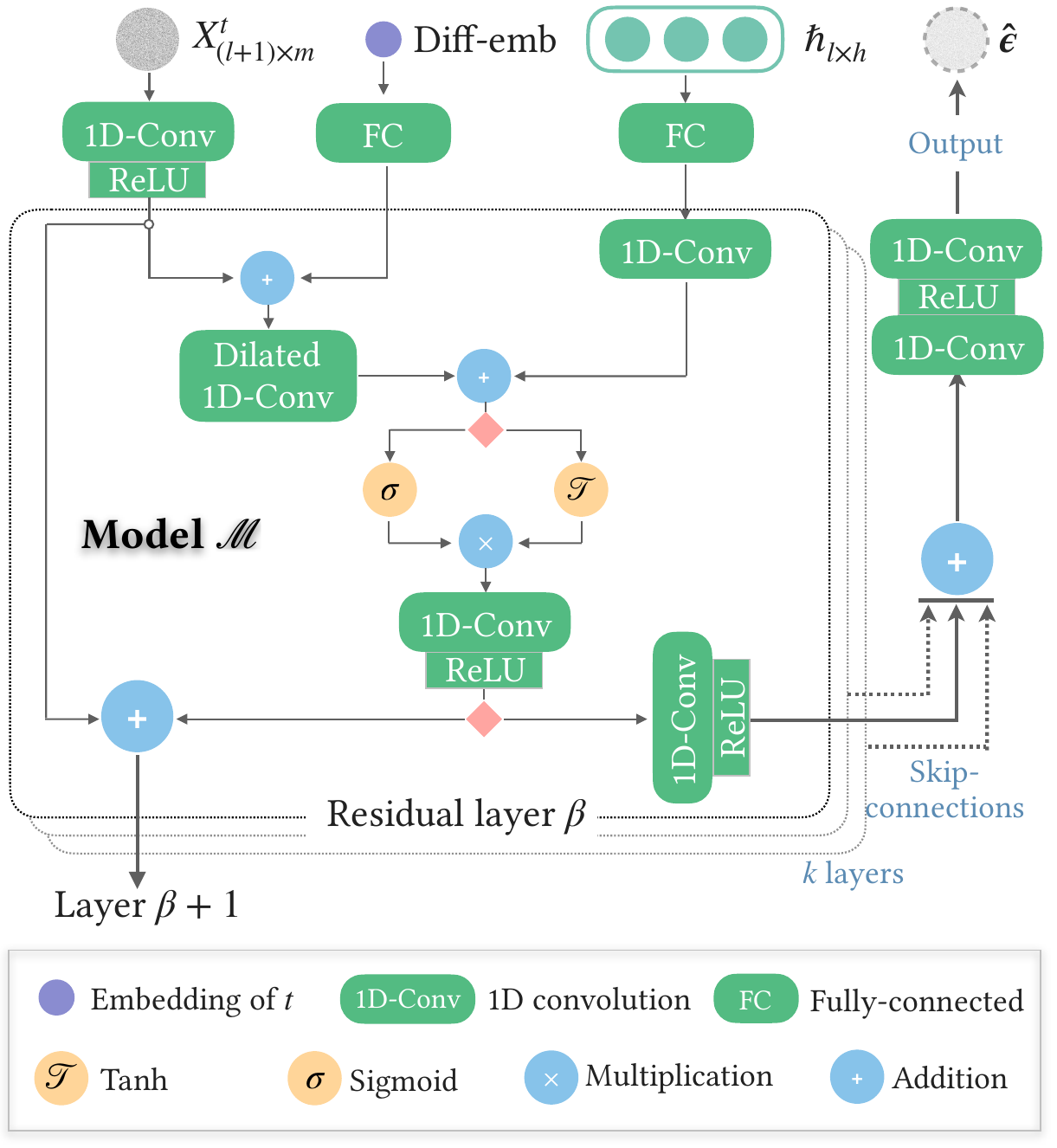}}
    \vspace{-0.1in}
    \caption{The architecture of model $\mathcal{M}$ fitting $\boldsymbol{\epsilon_\theta}(\cdot)$, where Diff-emb is the embedding of the step's index $t$.}
    \label{fig:modelM}
 \vspace{-0.1in}
\end{figure}

We transformed the 2D network of PixelCNN into 1D for our application, denoted by model $\mathcal{M}$, whose architecture is illustrated in Figure~\ref{fig:modelM}.
Model $\mathcal{M}$ consists of $k$ residual layers with skip-connections, each with multiple 1D convolution operators, a dilated 1D convolution operator~\cite{dilated}, and a gated activation unit that is calculated by:
\begin{equation}
    Y_{\text{output}} = \text{tanh}(W_1 \circledast X_{\text{input}}) \odot \sigma(W_2 \circledast X_{\text{input}})
\end{equation}
where $W_1$, $W_2$ are learnable parameters; $\circledast$ is the convolution operator, and $\odot$ is the element-wise product; $\sigma$ and $\text{tanh}$ are sigmoid and tanh functions, respectively.
In addition, we apply the sinusoidal position embedding leveraged in Transformer~\cite{Attn} to embed the diffusion step's index $t$.

\begin{table*}[htb]
    \centering
    \caption{Descriptions of some anomaly-indicating candidate features for general cloud services.}
    \begin{adjustbox}{max width=1.99\columnwidth}
    \begin{threeparttable}
        \begin{tabular}{c|c|c|c}
            \Xhline{1.5pt} 
            \xrowht[()]{12pt}
            \textbf{Category} & \textbf{Feature}               & \textbf{Description}                                                                                        & \textbf{Manifestation}                        \\
            \Xhline{1pt}
            \multirow{4}{*}{Point}
                              & (abs-)Min/Max                  & (absolute) extremums of computing resources (e.g., CPU, memory)                                             & alarming peaks                                         \\
            \Xcline{2-4}{0.3pt}
                              & ZeroCount                      & the number of points with values being zero                                                                 & dead process                                           \\
            \Xcline{2-4}{0.3pt}
                              & SpeDayCount                    & the number of holidays or festivals inside an observed window                                               & expected peaks                                         \\
            \Xcline{2-4}{0.3pt}
                              & OverZCount                     & the number of points with the modified Z-score~\cite{ModifiledZ} larger than 3.5                            & short-lived spikes                                     \\
            \Xhline{1pt}

            \multirow{4}{*}{\makecell[c]{Frequency                                                                                                                                                                                    \\ domain}}
                              & FC                             & Fourier coefficients of performance metrics                                                                 & \multirow{3}{*}{\makecell[c]{voilent fluctuations;     \\ jitters}}\\
            \Xcline{2-3}{0.3pt}
                              & FTP-param                      & parameters (centroid, variance, etc.) of absolute Fourier transform spectrum                                &                                                        \\
            \Xcline{2-3}{0.3pt}
                              & CPSD                           & cross power spectral density between metrics of the same aspect\tnote{$\dagger$}                            &                                                        \\
            \Xcline{2-4}{0.3pt}
                              & SK                             & time-varying spectral kurtosis~\cite{TVSK} of IO Bytes                                                      & nonstationary regime                                   \\
            \Xhline{1pt}

            \multirow{5}{*}{Trend}
                              & \multirow{2}{*}{LLS-param}     & \multirow{2}{*}{\makecell[c]{calculate the linear least-squares regression over the observed window                                                                  \\ and obtain the slope, intercept, standard error, etc.}} & \multirow{5}{*}{\makecell[c]{horizontal/up-/down- \\ trend; linearity;}}\\
                              &                                &                                                                                                             &                                                        \\
            \Xcline{2-3}{0.3pt}
                              & \multirow{2}{*}{LLS-agg-param} & \multirow{2}{*}{\makecell[c]{calculate the linear least-squares regression over rolling sub-sequences in an                                                          \\ observed window and obtain the mean of slopes, intercepts, standard errors, etc.}}&\\
                              &                                &                                                                                                             &                                                        \\
            \Xcline{2-3}{0.3pt}
                              & c3                             & c3 statistics~\cite{c3} of computing resources measuring the non-linearity                                  &                                                        \\
            \Xhline{1pt}

            \multirow{4}{*}{\makecell[c]{Temporal                                                                                                                                                                                     \\ dependencies}}
                              & ACF-$\mu$/-$\sigma$            & the mean and the variance over the autocorrelation for different lags                                       & \multirow{2}{*}{\makecell[c] unpredictable volatility} \\
            \Xcline{2-3}{0.3pt}
                              & PACF-$\mu$/-$\sigma$           & the mean and the variance over the partial autocorrelation for different lags                               &                                                        \\
            \Xcline{2-4}{0.3pt}
                              & margin-$\Sigma$                & the sum of changes between every two neighboring points of metrics                                           & \multirow{2}{*}{\makecell[c]{sudden rise               \\and fall}} \\
            \Xcline{2-3}{0.3pt}
                              & (abs)-mar-Min/Max              & (absolute) extremums of marginal change of computing resources                                              &                                                        \\
            \Xhline{1pt}

            \multirow{3}{*}{Distribution}
                              & std                            & standard deviation                                                                                          & \multirow{3}{*}{\makecell[c]{concept shift,            \\ staircases}}\\
            \Xcline{2-3}{0.3pt}
                              & skew, kurt                     & skewness and kurtosis of both single-series metrics and joint multi-variate metrics                         &                                                        \\
            \Xcline{2-3}{0.3pt}
                              & $q$-quantiles                  & the quantile of 10\%, 50\%, 90\%, and the anomaly ratio (if known)                                          &                                                        \\
            \Xhline{1pt}

            \multirow{4}{*}{Cross-series}
                              & CID                            & the complexity-invariant distance~\cite{CID} between metrics of the same aspect                             & complexity                                             \\
            \Xcline{2-4}{0.3pt}
                              & corr                           & Pearson correlations of metrics between the same and different aspects                                      & \multirow{3}{*}{\makecell[c]{cross-metric and          \\ cross-aspect relations}}\\
            \Xcline{2-3}{0.3pt}
                              & TLCC                           & time lag cross-correlations of metrics between the same and different aspects                               &                                                        \\
            \Xcline{2-3}{0.3pt}
                              & MI                             & pointwise mutual information~\cite{MI} of metrics between the same and different aspects                    &                                                        \\
            \Xhline{1.5pt}
        \end{tabular}\label{tab:feat}
        \begin{tablenotes}
            \footnotesize
            % \item[*] The potential anomalies, unexpected signal behaviors, or characteristics can be reflected in the corresponding features.
            \item[$\dagger$] Metrics reflecting diverse aspects may tend to behave differently, e.g., the disk usage is steady, while the CPU usage can fluctuate dramatically without anomalies~\cite{Hades}.
        \end{tablenotes}
    \end{threeparttable}
    \end{adjustbox}
    \vspace{-0.1in}
\end{table*}

\subsection{Faster-than-real-time Detection}
Directly combining an advanced data-driven real-time anomaly detector (usually deep learning-based) with our forecaster can cause practicality, interpretability, and adaptability issues.
First, overly complex models tend to be impractical for deployment by largely increasing the computation burden.
Second, deep learning-based detectors are often black-box and difficult to interpret so as to lose the trust of operation engineers.
Third, numerical anomalies may not always equate to operational anomalies in real-world services, making fully data-driven approaches unadaptable in some cases.
For example, an inactive service that constantly uses small memory (e.g., 0.1\%) and suddenly occupies zero memory may not trigger an alarm in completely data-driven approaches, as there are no spikes, jitters, or other common abnormal behaviors.
However, this is an undoubted anomaly because low and no memory usage are qualitatively different.
To address these issues, we propose a visible and intuitive anomaly detection algorithm.
We define a set of anomaly-indicating features of performance metrics and leverage a tree-based model to achieve the final anticipation.

\subsubsection{Anomaly-indicting Feature Extraction}
This step extracts meaningful features from performance metrics.
We systematically identify a set of candidate features that can indicate common anomalies in cloud services based on our expertise and collaboration with industry partners. 
Representative features are presented in Table~\ref{tab:feat}, while additional feature definitions and the calculation code are also available in our online repository. This can also contribute to real-time anomaly detection in the software engineering domain.
These features cover typical patterns, such as spikes, jitters, and peaks, and incorporate system understanding and historical failure causes. For instance, a rise in the frequency of virtual memory usage may indicate more virtual page errors. Some features also consider human activities, such as holidays, which may impact network throughput for a ticket service.
Besides our defined features, we encourage developers to extend and enhance \tool by customizing the anomaly-indicating features for specific cloud service scenarios.
For example, a weather report service may experience a sudden surge in page views when a hurricane approaches, leading to an increase in network traffic. In such a case, developers may not want frequent alerts triggered by such surges, as fluctuations are expected.

\subsubsection{Feature Selection}
This step aims to identify the optimal subset of features by eliminating irrelevant or redundant features, thereby reducing the number of features to improve model accuracy and reduce execution time. Moreover, selecting genuinely relevant features simplifies the model and aids in understanding the data.
We select features based on the importance scores of features calculated by Xgboost on a small set with annotations (i.e., the validation set).
However, manual annotation is time-consuming and sometimes infeasible. 
For such cases, we can use the Pearson correlation coefficient or mutual information to calculate the relation between features and remove those strongly related features. 
Note that all existing feature selection approaches can be directly leveraged in this step, and expertise-dependent selection may perform better in the real world.

\subsubsection{Anomaly Anticipation}
The final step performs anomaly detection on the concatenation of observations and forecasts in an unsupervised manner.
Particularly, \tool adopts isolation forest~\cite{IF} to detect anomalies based on the extracted features.
Isolation forest can separate anomalous instances from the rest. It is prevalent in practice due to its high effectiveness, efficiency (with a linear time complexity), and intuitive interpretability, which is important for winning the trust of operation engineers.

Specifically, we first construct an isolation forest model with observations in the training set. The training data $X_{[1:N]}={x_{[1:l]}}{n=1}^N$ consists of $N$ data windows, where each window $x{[1:l]} \in \mathbb{R}^{m\times l}$ has $m$ metrics with length $l$. 
Each isolation tree (denoted by $iTree$) inside the forest samples a subset of the inputted data with the size of $\psi$ (a hyperparameter) and recursively constructs leaf nodes based on the values of the data attributes until the tree height reaches a pre-defined threshold, or all sampled data are used.

\begin{algorithm}[htbp]
    \small
    \caption{Incrementally training isolation forest.}
    \label{algo:detection}
    \LinesNumbered
    \KwIn{$X^{\text{cat}}_{[1:N]}$, $\gamma$, $\psi$, $F_{pre}$ - previously trained forest}
    \KwOut{A new forest $F$ consisting of $\gamma$ trees and $F_{pre}$}
    \textbf{Initialize} $F$ \\
    $i \leftarrow 1$
    \While{$i \leq \gamma$}{
      $X' \leftarrow sample(X^{\text{cat}}_{[1:N]}, \psi)$  \\
      $X'_{iso} \leftarrow F_{pre}(X')$ // Keep the samples ``isolated'' by $F_{pre}$\\
      $F \leftarrow F \cup iTree(X_{iso})$ \\
    }
    \Return{$F$}
\end{algorithm}

Assuming that the trained isolation forest contains $\gamma$ trees, we apply incremental learning to further train the model with forecasts, whose process is described in algorithm~\ref{algo:detection}.
The forecasted results $\hat{X}{[1:N]}={\hat{x}{[l+1:l+s]}}{n=1}^N$ are concatenated with the observed data, denoted by $X^{\text{cat}}{[1:N]} = {[x_{[s:l]}; \hat{x}{[l+1:l+s]}}{n=1}^N$. 
The concatenated data are then fed into the established forest. 
If a concatenation is isolated, we then remove it from the existing forest and build up a new isolation tree until the number of trees reaches the pre-defined threshold $\gamma$.
Finally, $2\gamma$ trees are created in the isolation forest.
In this way, only the extremely abnormal concatenated samples are isolated. 
The idea is based on the fact that anomalies are rare and most of the concatenated samples should be normal, even though there are slight differences between concatenated ones and the original observations $X_{[1:N]}$.

\section{Evaluation}\label{sec:experiment}
We evaluate \tool by answering three research questions:
\begin{itemize}[leftmargin=10pt, topsep=0pt]
\item RQ1: How effective is \tool in anomaly anticipation?
\item RQ2: How effective is the forcaster of \tool?
\item RQ3: How much time can \tool advance anomaly alarm?
\end{itemize}

\subsection{Experiment Settings}
\subsubsection{Comparative Approaches}
As \tool is the first work to address anomaly anticipation, we have to compare it with real-time anomaly detectors. 
Following the most recent anomaly detection papers~\cite{Hades, adsketch}, we choose eight state-of-the-art baselines: Dount~\cite{dount}, SR-CNN~\cite{srcnn}, Adsketch~\cite{adsketch}, Telemanom~\cite{telemanom}, LSTM-VAE~\cite{lstm-vae}, MTAD-GAT~\cite{mtad}, DAGMM~\cite{dagmm}, and OmniAnomaly~\cite{OmniAnomaly}.
Note that \tool anticipates anomalies in advance, whereas the competitors issue alerts after their occurrences.
We also evaluate \tool's detector individually by removing the forecaster and only retaining the detector before incremental training to derive its real-time version, \tool-rt, trained and tested on observations as baselines do.

As far as we know, no existing studies directly target performance metric forecasting. Thus, we compare \tool's forecaster with general-purpose deterministic baselines (GRU~\cite{gru}, TCN~\cite{TCN}, and Transformer~\cite{Attn}) and advanced probabilistic generation-based methods for multivariate time series (DeepVAR~\cite{DeepVAR}, GRU-MAF, and Transformer-MAF~\cite{condFlow}). 
The latter category regards the mean of output as the final forecast.

\subsubsection{Datasets}
We use three wildly used datasets containing complex anomaly patterns. 
Table~\ref{tab:dataset} summarizes the statistics of these datasets, where \#MetricNum denotes the number of metrics in the dataset, and \#MetricLen Avg. denotes the average number of sampled points of each metric. 
All of them are publicly available with brief introductions as follows.
\begin{itemize}[leftmargin=*, topsep=10pt]
    \item \textit{AIOps18}~\cite{AIOps18} is a real-world dataset consisting of business and performance metrics (inside this paper's scope) collected from web services of a large-scale IT company. The metric interval is either one minute or five minutes. 
    \item \textit{Hades} is a recently released dataset for anomaly detection~\cite{Hades}, containing multivariate performance metrics and logs collected from Apache Spark with annotations. It covers various workloads and 21 typical types of faults. The metrics are all equally spaced one second apart.
    \item \textit{Yahoo!S5} is a benchmark dataset for metric-based anomaly detection~\cite{yahoo}. We only consider its real-world metrics sampled every hour, whose anomalies are labeled by hand. 
\end{itemize}
AIOps18 and Hades are multivariate, while Yahoo!S5 is univariate. 
We choose these different datasets because the real-world needs of engineers may vary from holistic detection over multiple metrics to focusing on a single essential metric.

\begin{table}[htb]
\vspace{-0.1in}
\small
\centering
\caption{Dataset statistics.}
\vspace{-0.1in}
\begin{tabular}{cccc}
\toprule
Dataset & \#MetricNum & \#MetricLen Avg. & Anomaly Ratio  \\
\midrule
AIOps18 & 29 & 204238.38 & 2.26\% \\
Hades & 11 & 64,422 & 24.47\% \\
Yahoo!S5 & 67 & 1415.91 & 1.76\% \\
\bottomrule
\end{tabular}\label{tab:dataset}
\vspace{-0.1in}
\end{table}

The used datasets are split based on their collection time: the first 60\% is for training, and the subsequent 10\% and 30\% are for validation and testing, respectively. This splitting guarantees no data leakage with respect to time.
Moreover, all methods, except Adsketch, are unsupervised, so only Adsketch requires normal data for training. 
We first follow Adsketch~\cite{adsketch}'s original paper to obtain the training and test sets for AIOps18 and Yahoo!S5.
Then we follow Hades's paper~\cite{Hades} to train Adsketch with fault-free data.
The validation and testing sets are randomly sampled from the previously split datasets to make the final ratio 6:1:3. 

\subsubsection{Evaluation Measurements}
The final results of anomaly anticipation are binary (normal or not), so we adopt the widely-used evaluation measurements of binary classification to gauge \tool:
\begin{equation}
    \textit{Rec}=\frac{\textit{TP}}{\textit{TP}+\textit{FN}}, \textit{Pre}=\frac{\textit{TP}}{\textit{TP}+\textit{FP}}, \textit{F1}=\frac{2\cdot \textit{TP}}{2\cdot \textit{TP}+\textit{FN}+\textit{TP}}
\end{equation}
where \textit{TP} is the number of correctly detected abnormal samples; 
\textit{FP} is the number of normal samples incorrectly triggering alarms;
\textit{FN} is the number of undetected abnormal samples.
Performance metrics in one observation window form a univariate/multivariate sample.

Additionally, we evaluate our forecaster separately with two measurements:
mean absolute error and symmetric mean absolute percentage error:
\begin{equation}
    \textit{MAE}=\frac{1}{N}\sum_{i=1}^N (\hat{y}_i-y_i)^2),
    \textit{sMAPE}=\frac{1}{N}\sum_{i=1}^N \frac{|\hat{y}_i-y_i|}{(|\hat{y}_i|+|y_i|)/2})
\end{equation}
where $y_i$ and $\hat{y}_i$ are the real and forecasted values at time $i$, respectively; $N$ denotes the number of samples.
$y_i$ is a float if the input is univariate, and otherwise, for a $m$-variate input, $y_i \in \mathbb{R}^m$ is a vector.
Absolute error-based measurements are preferred because we intend to encourage aggressive forecasting but squared error-based measurements, such as mean squared error (MSE), tend to punish extreme forecasted values.
Moreover, MAE is simple to understand, engaging its prevalence in practice; sMAPE is expressed as a percentage, which is scale-independent and suitable for comparing forecasts on different scales. 

\begin{table*}[htb]
\centering
    \vspace{-0.1in}
    \caption{Overall performance comparison (\%)*.}
    \vspace{-0.1in}
        \begin{threeparttable}
        %\begin{adjustbox}{max width=\columnwidth*3}
        \begin{tabular}{*{11}{c}|*{3}{c}}
        \toprule
        \multirow{2}*{\textbf{Mode}} & \multirow{2}*{\textbf{Methods}} & \multicolumn{3}{c}{AIOps18\tnote{$\dagger$}} & \multicolumn{3}{c}{Hades} & \multicolumn{3}{c}{Yahoo! S5} & \multicolumn{3}{c}{\textbf{Average}}\\
        \cmidrule(lr){3-5}\cmidrule(lr){6-8}\cmidrule(lr){9-11}\cmidrule(lr){12-14}
        
            &&\textit{F1} & \textit{Rec} & \textit{Pre}
            &\textit{F1} & \textit{Rec} & \textit{Pre}
            &\textit{F1} & \textit{Rec} & \textit{Pre}
            &\textit{F1} & \textit{Rec} & \textit{Pre}\\
            \cmidrule{1-14}\morecmidrules\cmidrule{1-14}
          \multirow{9}*{\makecell[c]{\textbf{Real-} \\ \textbf{time}}}
          & Dount       & 36.60 & 43.06 & 31.82 & 49.17 & 47.49 & 50.97 
                        & 58.30 & 65.77 & 52.36 & 48.02 & 52.11 & 45.05\\
          & SR-CNN      & 44.81 & \textbf{71.91} & 32.54 & 34.25 & 61.43 & 23.74 
                        & 41.06 & 61.81 & 30.74 & 40.04 & 65.05 & 29.01\\
          & Adsketch    & \underline{64.82} & 64.28 & 65.37 & 65.35 & 57.47 & 75.73 
                        & 58.08 & 67.28 & 51.09 & 62.75 & 63.01 & 64.06\\
          & Telemanom   & 49.49 & 60.10 & 42.06 & 46.75 & 66.29 & 36.10 
                        & 54.10 & \textbf{77.43} & 41.57 & 50.11 & 67.94 & 39.91\\
          & LSTM-VAE    & 46.35 & 54.57 & 40.29 & 36.89 & 69.07 & 25.16 
                        & 62.77 & 63.35 & 62.20 & 48.67 & 62.33 & 42.55\\
          & MTAD-GAT    & 37.85 & 46.24 & 32.04 & 56.90 & 55.40 & 58.48 
                        & 35.62 & 31.86 & 40.38 & 43.46 & 44.50 & 43.63\\
          & DAGMM       & 53.52 & 58.08 & 49.63 & 62.10 & 55.62 & 70.29 
                        & 57.33 & 51.70 & 64.33 & 57.65 & 55.13 & 61.42\\
          & OmniAnomaly & 57.40 & 66.82 & 50.31 & 68.17 & 78.81 & 60.06
                        & 53.13 & 76.75 & 40.63 & 59.57 & 74.13 & 50.33\\
          \cmidrule(lr){2-14}
          %\cdashline{2-14}[0.8pt/2pt] 
          & \textbf{\tool-rt} & \textbf{66.75} & 64.12 & \textbf{69.60}
                        & \textbf{85.30} & 84.35 & \textbf{86.28} & \textbf{72.28} & 74.65 & 70.06 & \textbf{74.78} & \textbf{74.37} & \textbf{75.31} \\
          \midrule
          %\cline{1-1}
          \textbf{FTRT} & \textbf{\tool} & 63.78 & 58.94 & 69.48
                        & 82.07 & \textbf{88.77} & 76.31 & 70.31 & 69.15 & \textbf{71.51} & 72.05 & 72.29 & 72.43\\
          \bottomrule
        \end{tabular}
        %\end{adjustbox}

        \begin{tablenotes}
        \footnotesize
        \item[*] Each value is averaged over three independent runs with different random seeds under the optimal hyperparameters.
        \item[$\dagger$] In AIOps18, only metrics collected at the intersecting timestamps are considered, resulting in a 16-variate time series.
        %\item[$\ddagger$] FTRT: faster-than-real-time, i.e., the anticipating mode.
      \end{tablenotes}
      
\vspace{-0.1in}
\end{threeparttable}\label{tab:overall-exp}

\vspace{-0.1in}
\end{table*}

%iForest 0.4013 0.4153 0.3882 for aiops18
%iForest 0.437 0.597 0.713 
% MTAD 0.7124 0.5540 0.9976 for Hades 

\subsubsection{Implementations and Hyperparameters}
Our implementation is based on Python 3.8 with open-source packages such as Scikit-learn, Pytorch, TSfresh~\cite{TSfresh}, and Gluonts~\cite{gluonts}.
We directly deploy the original open-source implementations of all compared methods. 
In particular, we converted point-wise detection results into segment-level results for fairness by considering the entire window as abnormal once an anomalous data point is detected, which is presented by~\cite{dount} and applied in many baselines' original papers~\cite{dount, mtad, dagmm, OmniAnomaly}.
All of the experiments are performed on three NVIDIA GeForce GTX 1080 GPUs. We use an open-source toolkit called NNI~\cite{nni} to optimize the hyperparameters of all approaches automatically, and the best validation \textit{F1} corresponds to the optimal hyperparameter combination.
 
As for hyperparameters in \tool, the hidden size of metric embeddings is 64, and the number of sampling time steps is 100. We adopt an early-stop mechanism in the forecaster, and the maximum epoch is 50. In each epoch, 100 batches are randomly selected, and the batch size is 256. We use the Adam optimizer with the initial learning rate set to 0.0005. 
To address the large number of anomaly-indicating features induced by high-dimensional metrics, we apply Xgboost~\cite{xgboost} on the validation set to filter features before detection. The detector hyperparameters are default values in Scikit-learn.

\subsection{RQ1: Overall Effectiveness of \tool}\label{sec:exp:RQ1}
From Table~\ref{tab:overall-exp}, we can conclude that \tool achieves comparable or even superior effectiveness as an FTRT anomaly anticipator compared to real-time detectors. 
\tool outperforms all other approaches (except \tool-rt) in terms of the major evaluation measurement \textit{F1}, with \textit{F1} scores being 15.15\% to 79.95\% higher than competitors on average.
\tool's \textit{F1} is only slightly lower than \tool-rt.
Note that it is reasonable and expected that \tool's detection is slightly poorer than its real-time version because the forecaster is inevitable to bring imperfections. 
The slight inferiority actually indicates \tool can make accurate forecasts and deliver results effectively in advance.
This finding is important as it suggests that \tool can be a viable approach for anticipating anomalies before they occur in cloud services without sacrificing detection accuracy.

In particular, the superiority of \tool is more significant in Hades and Yahoo!S5, with \textit{F1} being 20.30\% to 139.62\% higher than competitors for Hades and 12.01\% to 97.39\% higher for Yahoo!S5.
Moreover, \tool's \textit{F1} is only 1.60\% lower than the best competitor, Adsketch, for AIOps18.
Notably, Adsketch is a semi-supervised approach that requires normal data as training inputs, while \tool can resist noisy training data containing anomalies.
Furthermore, \tool shows superiority over Adsketch in AIOps18 and Hades, as Yahoo!S5 contains relatively simple data patterns with a low anomaly ratio, where noisy training data has a lesser impact on detection.

In terms of \textit{Pre}, \tool surpasses all other real-time methods by 13.07\%$\sim$149.67\% on average, indicating that it is effective in targeting anomalies. 
Some methods achieve higher \textit{Rec} scores on specific datasets, such as SR-CNN and OmniAnomaly on AIOps18 and Telemanom on Yahoo!S5.
However, they all perform an imbalance between \textit{Rec} and \textit{Pre}.
Note that AIOps18 is a complex dataset with diverse data patterns, making it difficult to model essential normal patterns, resulting in more false positives.
In the case of SR-CNN, Spectral Residual (SR) labeling may cause more false positives by regarding complicated data as abnormal, while OmniAnomaly's assumption of generalized Pareto distribution and Gaussian distribution may not hold true in real-world scenarios. 
Telemanom, on the other hand, predicts the next-time metric and can mistake the inaccurate prediction for anomalies, leading to low \textit{Pre}. 
Overall, \tool achieves a balance between \textit{Pre} and \textit{Rec}, making it a robust and effective anomaly anticipator for various real-world scenarios.

Moreover, \tool-rt outperforms all competitors by 19.17\%$\sim$86.76\% on averaged \textit{F1}.
Unlike other fully data-driven techniques, \tool-rt incorporates human expertise to design anomaly-indicating features, able to detect suspicious metric segments more in line with operation engineers' desire. 
In addition, the isolation forest used in \tool-rt fits the most concentrated regions of the training data, regardless of some anomalies, thereby being robust to training noise that wildly exists in real-world data. Besides, it does not assume a priori distributions, enabling \tool-rt to handle different data distributions in practice.

\begin{mybox}
  \small
  \textbf{A1:} \tool, able to provide the FTRT anomaly anticipation relying on forecasts, performs as well as or better than state-of-the-art real-time detectors based on real observations. 
\end{mybox}

\subsection{RQ2: Effectiveness of Forecasting}
Table~\ref{tab:exp:pred} presents the evaluation on \tool's forecaster(\tool-$\mathcal{F}$), demonstrating that \tool-$\mathcal{F}$ outperforms baselines significantly, particularly on complex datasets (AIOps18 and Hades), reducing \textit{MSE} by 44.73\%$\sim$89.81\% and \textit{sMAPE} by 30.76\%$\sim$65.87\% on average.

\begin{table}[htbp]
\small
\centering
\caption{Comparison for performance metric forecasting.}
\vspace{-0.1in}
    \begin{threeparttable}
    \begin{adjustbox}{max width=\columnwidth}
    \begin{tabular}{l*{7}{c}}
    \toprule
    \multirow{2}*{\centering \textbf{Methods}} & \multicolumn{2}{c}{AIOps18} & \multicolumn{2}{c}{Hades} & \multicolumn{2}{c}{Yahoo!S5}\\
    \cmidrule(lr){2-3}\cmidrule(lr){4-5}\cmidrule(lr){6-7}
    
    &\textit{MSE}&\textit{sMAPE}&\textit{MSE}&\textit{sMAPE}&\textit{MSE}&\textit{sMAPE}\\
    \cmidrule{1-8}\morecmidrules\cmidrule{1-8}
    GRU      & 6.170 & 1.256 & 3.368 & 1.957 & 1.422 & 1.448\\
    TF*      & 5.627 & 1.400 & 5.628 & 1.492 & 1.717 & 1.443\\
    TCN      & 4.610 & 1.230 & 3.622 & 0.835 & 1.111 & 1.498\\
    %\hdashrule[0.5ex][x]{\linewidth}{0.5pt}{1.5mm}
    %\hdashline[2pt/5pt]
    DeepVAR  & 0.428 & 0.677 & 1.250 & 0.692 & 0.714 & 1.022\\
    GRU-MAF  & 2.607 & 1.451 & 6.739 & 1.959 & 1.180 & 1.439\\
    TF-MAF   & 3.235 & 1.470 & 2.091 & 1.677 & 1.226 & 1.505\\
    \midrule
    \textbf{\tool}-$\mathcal{F}$ & \textbf{0.298} & \textbf{0.566} & \textbf{0.597} & \textbf{0.487} & \textbf{0.426} & \textbf{0.602}\\
    \bottomrule
    \end{tabular}
     \end{adjustbox}
    \begin{tablenotes}
            \footnotesize
            \item[*] TF is Transformer.
        \end{tablenotes}
    \end{threeparttable}
\label{tab:exp:pred}
\end{table}

The success of \tool-$\mathcal{F}$ can be attributed to three reasons:
1) The diffusion model's flexibility allows it to forecast abnormal values exceeding the value limit of the contexts, resulting in better and more aggressive forecasting, especially anomalous segments.
In contrast, classical baselines tend to make conventional forecasting limited to the range of inputs by directly projecting the historically observed data into the output space, failing to forecast abnormal values with a low probability.
2) Unlike flow-based approaches that are constrained by the invertibility of a sequence of transformations, which limits the expressiveness of the model, our diffusion design can better model the predictive distribution.
This explains why MAF-based approaches perform better than traditional deep learning models but still suffer inaccuracy for complex data. 
Moreover, their failure to estimate the likelihood of out-of-distribution samples makes them even worse, since our data contain multiple anomalies falling off the assumed distributions.
3) \tool-$\mathcal{F}$ is more robust to input noise as it incorporates metric values through the likelihood term rather than directly modeling these values, which is adopted in DeepVAR.
\tool-$\mathcal{F}$ demonstrates the greatest improvement in Hades dataset, which contains more anomalous metric segments, with a 52.19\% reduction in \textit{MSE} and a 29.60\% reduction in \textit{sMAPE}.
These results highlight the significant contribution of \tool-$\mathcal{F}$ in improving accuracy for complex and anomalous datasets.

\begin{mybox}
  \small
  \textbf{A2:} \tool's forecaster performs effectively in metric forecasting, especially in anomalous contexts. Such accuracy enables \tool to attain effective anomaly anticipation results.
\end{mybox}

\subsection{RQ3: Time Advancing of \tool}
Table~\ref{tab:exp:time} presents the overhead of each phase in \tool's inference, including forecasting (\#ForeT), feature extraction (\#FeatT), and detection (\#DeteT), where \#ForeL denote the observation length and the forecast length, respectively.
\#ALen indicates how far in advance can \tool alarm an upcoming anomaly, and the advanced time equals the production of \#ALen and the metric sampling interval. For example, Yahoo!S5 is sampled every hour, and \tool can report anomalies 3 hours faster than real-time under our setting, though the potential has not been fully exploited.

\begin{table}[htb]
\small
\centering
\vspace{-0.1in}
\caption{Time consumption of \tool (unit: second).}
\vspace{-0.1in}
    \begin{tabular}{cccccc}
    \toprule
    Dataset  & \#ALen & \#PredT & \#FeatT & \#DeteT & Advanced\\
    \midrule
    AIOps18  & 5      & 3.031   & 1.320   & 0.035   & 4.386 \\ %3.170 avg.
    Hades    & 3      & 1.922   & 0.976   & 0.036   & 2.934 \\
    Yahoo!S5 & 3      & 1.915   & 0.238   & 0.036   & 2.189 \\
    \bottomrule
\end{tabular}\label{tab:exp:time}
\end{table}
% Dount
% SRCNN
% Adsketch
% Tele
% LSTM-VAE
% MTAD-GAT
% DAGMM
% Omni

The experimental results show that \tool can effectively anticipate upcoming anomalies 3$\sim$5 time points in advance, saving significant time for downstream analysis.
This is because the interval of metric sampling is usually on the order of minutes or hours in practice, while the anticipation time is just seconds, almost negligible in contrast.
In addition, the anticipation paradigm of \tool has the potential to prevent anomalies and even serious failures before their occurrence by providing forecasts for downstream automated analysis.

\begin{mybox}
  \small
    \textbf{A3:} \tool can anticipate anomalies several minutes or hours in advance, with only a few seconds needed for inference, thus imposing negligible overhead on the system. This means that \tool saves a lot of time for downstream analysis and has the potential to prevent anomalies and more serious failures.
\end{mybox}

\subsection{Successful metric forecasts and feature extraction}\label{sec:case-study}
\begin{figure}[htbp]
\vspace{-0.1in}
\centering
\subfigure[AIOps18: Metric ``8723f0fb-eaef-32e6-b372-6034c9c04b80'']{
    \includegraphics[width=0.9\linewidth]{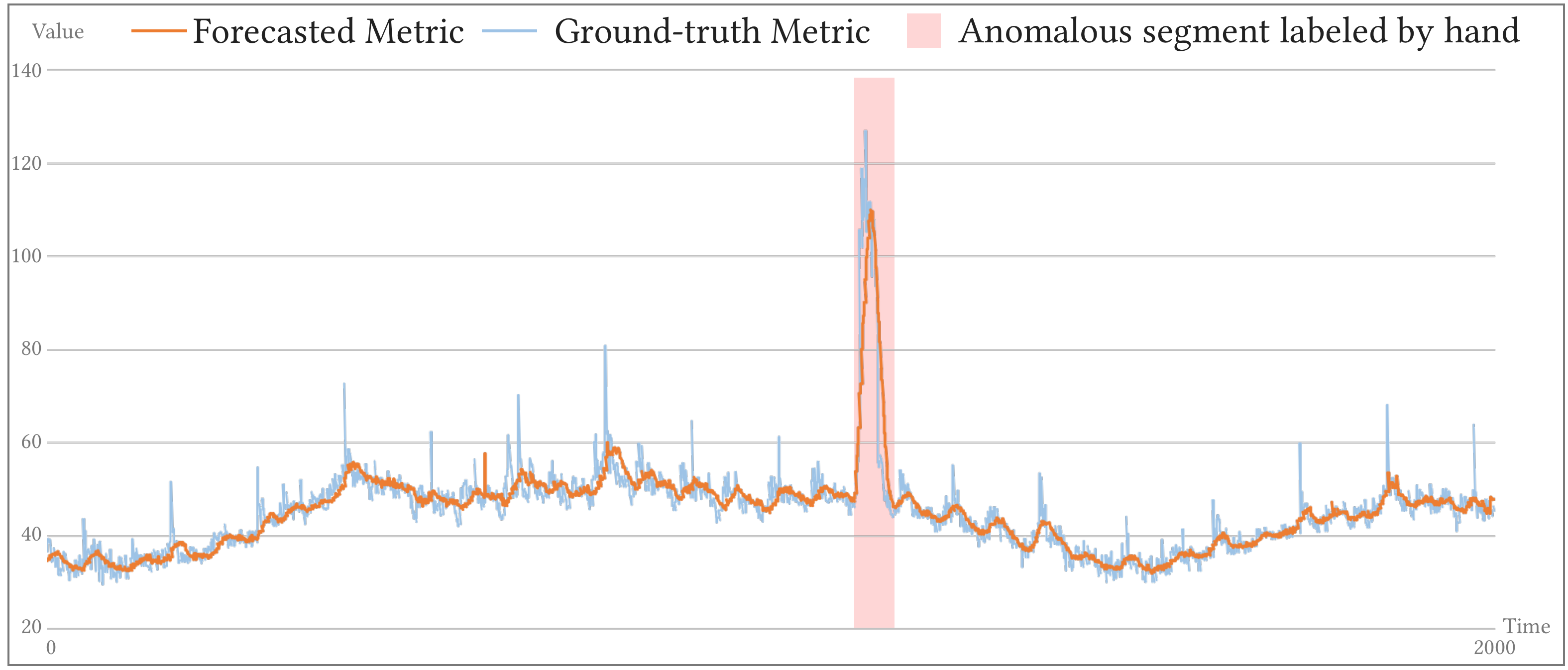}
}
\subfigure[Hades: Metric ``CPU iowait'']{
    \includegraphics[width=0.9\linewidth]{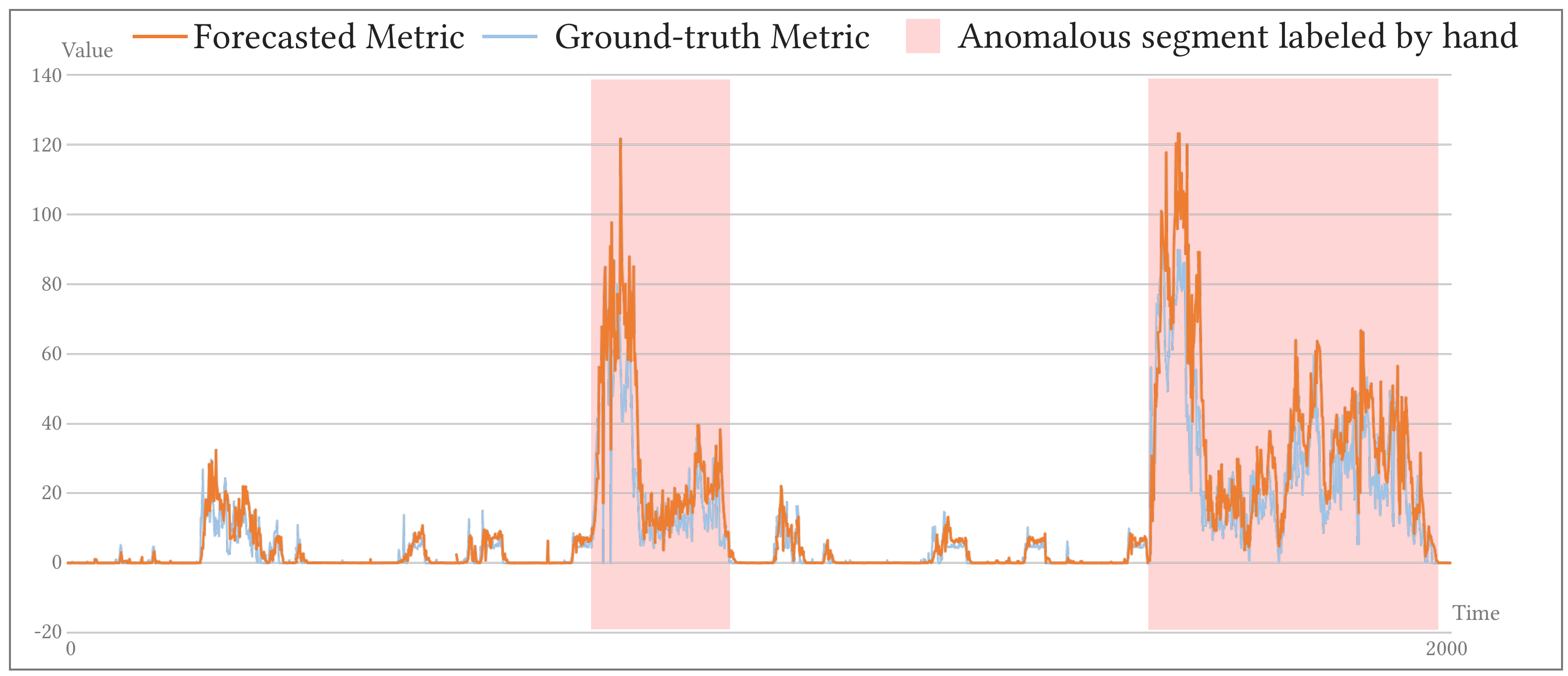}
}
\vspace{-0.1in}
\caption{Successful cases of \tool in forecasting metrics with anomalous segments on AIOps18 and Hades.}
\vspace{-0.2in}
\label{fig:pred-case}
\end{figure}

We present cases highlighting \tool's success in forecasting abnormal metrics and utilizing anomaly-indicating features.
Specifically, Figure~\ref{fig:pred-case} displays that \tool can accurately forecast performance metrics, even in abnormal contexts. 
Though \tool may not be able to forecast the exact same values, and it is almost impossible to do so, it successfully forecasts the trend of suspicious plunges or drops, which implies its ability to issue effective warnings ahead of anomalies' occurrence. 

\begin{figure}[htbp]
\vspace{-0.1in}
\centering
\subfigure[Real-17]{
\includegraphics[width=0.45\linewidth]{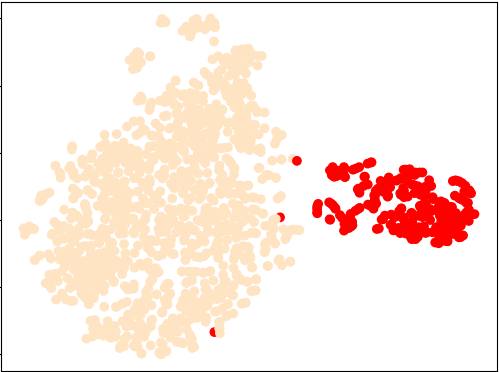}
}
\subfigure[Real-19]{
\includegraphics[width=0.45\linewidth]{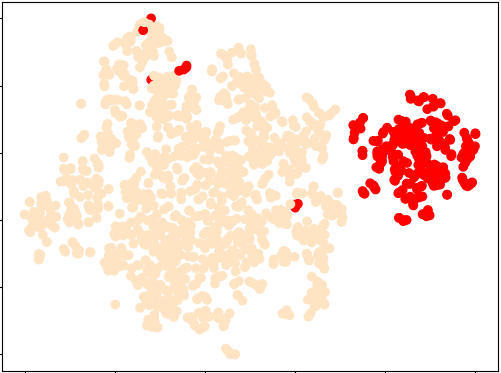}
}
\subfigure[Real-22]{
\includegraphics[width=0.45\linewidth]{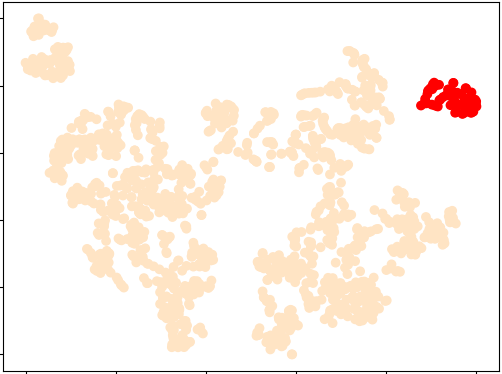}
}
\subfigure[Real-42]{
\includegraphics[width=0.45\linewidth]{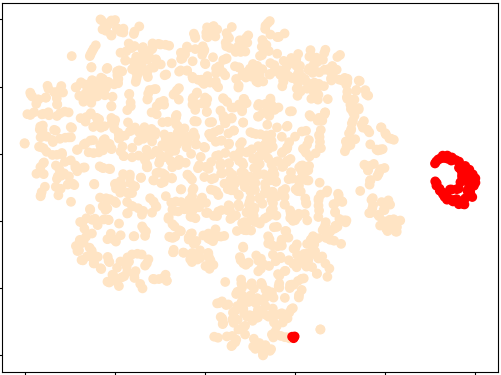}
}
\vspace{-0.1in}
\caption{Distributions of the anomaly-indicating features of four metrics in Yahoo!S5,  wherein the \bisque{bisque} points represent normal samples and the \red{red} points represent abnormal samples.}
\vspace{-0.1in}
\label{fig:feat-case}
\end{figure}

Moreover, we find that using the defined anomaly-indicating features enables the effective isolation of anomalies from normal samples. 
To illustrate this, Figure~\ref{fig:feat-case} showcases the feature distributions of four metrics (named by Real-*) from Yahoo!S5. 
The figure is constructed by t-SNE~\cite{t-SNE} with principal component analysis as the inner dimension reduction technique to project the features into a 2D space.  
Notably, the extracted anomaly-indicating features directly facilitate easy discrimination between anomalies and normal samples. Thus, we posit that properly designed features contribute to the success of \tool in anomaly detection.

\section{Discussion}\label{sec:discuss}
\subsection{Limitations}\label{sec:limitation}
\noindent\textit{Limited generalizability beyond cloud service performance.}
\tool's anomaly anticipation is tailored to cloud-service performance metrics with predictable anomalies due to their well-explored temporal relations and similar underlying infrastructures.
Its effectiveness for other types of time series or external anomalies is uncertain as other anomalies beyond cloud-service performance may be unpredictable.

\noindent\textit{Pre-defined advancing time.}
Engineers must decide the time horizon for anomaly anticipation at the very beginning, and \tool cannot adapt to changes once trained. This also leads to an insufficient explosion of \tool's potential since it may also work with an extended horizon, but the effectiveness under such a setting is unknown.
Additionally, one trained model can not fit all services, though a one-size-fits-all solution is desirable to handle diverse situations in practice.

\noindent\textit{Limited applicability in high-frequency sampling.}
\tool only needs a few seconds for inference and thereby is efficient for most real-world cloud systems with a moderate sampling frequency, allowing anticipation a few minutes or hours ahead. 
Though most systems apply minutely or hourly sampling frequency due to computational and storage resource limitations, with extremely high sampling frequency, \tool's forecasting inference may become slower than real-time observation, making it impractical for anomaly anticipation.

\subsection{Threats to Validity}\label{sec:threat}
\noindent\textit{Internal.}
Despite \tool's success in anticipating anomalies, there is room for improvement in forecasting spikes.
The design of the learning rate decay function and loss function may still need refinement to prevent the coverage at a local optimum or overly conservative forecasting. 
To alleviate this threat, we choose the smooth L1 loss instead of the most commonly used L2 loss to encourage aggressive forecasting, whose usefulness is demonstrated in the encouraging experimental results and case studies.
On the other hand, occasional unpredictable spikes may not significantly threaten the validity of \tool as it will issue an alert at the anomaly's onset if the anomaly has a lasting impact and thereby forms an anomalous segment. Otherwise, real-time detection is acceptable if the spike's cause only incurs instantaneous effects.

\noindent\textit{External.}
The effectiveness of \tool on other datasets is yet unknown. 
However, we carefully select three representative datasets for evaluation.
AIOps18 and Yahoo!S5 are real-world collections from big companies, widely used in existing comparable studies.
Hades covers multiple workloads and typical fault types derived from a large-scale company, thereby supporting its representativeness.

\section{Related Work}\label{sec:review}
\subsection{Early Anomaly Detection}
Early anomaly detection is a challenging and emerging problem in various domains, including cloud computing.
Existing studies (including but not limited to the cloud computing domain) focus on mining the relationships between observed metrics and future status. 
Some methods directly predict the binary status (normal or abnormal) in the future~\cite{Robust09, eWarn, PreGAN, NTAM, SMART, CDEF}, while others expand real-time binary classification (normal or abnormal currently) into a triple-state classification by adding the state of anomalies about to occur~\cite{BlackBox-AP, Hosting10}.
Despite their success, these methods by means of supervised learning require large amounts of high-quality labels and suffer from imbalanced data, making them difficult to apply in reality. 
Practically, the number of abnormal or alerting data samples is often much smaller than that of the normal ones.
Additionally, they struggle to generate high-quality intermediate forecasts for downstream analysis after anomaly detection, limiting their potential to take full advantage of early detection and steer services from failure.

\subsection{Real-time Anomaly Detection}
Existing anomaly detection approaches fall into two categories: statistical or machine learning-based~\cite{ARMA, adsketch, LeSiNN} and deep learning-based~\cite{dount, OmniAnomaly, mtad, dagmm, lstm-vae, telemanom, HTMBN, Eadro, Hades, SCWarn}.
Deep learning-based approaches model temporal dependencies and cross-metric relationships, and are mostly reconstruction-based or forecasting-based. 
Reconstruction-based approaches reconstruct the probability distribution of normal data via the family of auto-encoders~\cite{dount, OmniAnomaly, lstm-vae}.
Forecasting-based approaches~\cite{telemanom, HTMBN} assume that the forecasted values conform to the distribution of the observations. This assumption is both plausible and flawed, especially when it comes to complex distributions. 
MTAD-GAT~\cite{mtad} combines both, improving the detection accuracy while dramatically increasing the approach's complexity.
Moreover, some studies incorporate multi-modal information like logs~\cite{Hades, SCWarn, Eadro}, and promising detection performance has been achieved.
However, traditional real-time anomaly detectors are reactive and can only alarm anomalies after occurrences, which may be a bit late. Hence, we propose a novel anomaly anticipation approach to issue anomalies in advance, allowing for pre-anomaly intervention and saving time for downstream analysis.
\section{Conclusion}\label{sec:conclusion}
This paper proposes and formulates the paradigm anomaly anticipation for improving cloud service reliability for the first time, which consists of two stages: performance metric forecasting and anomaly detection on forecasts. 
We propose \tool, the first framework to address this problem. 
\tool presents a novel conditional denoising diffusion model and defines anomaly-indicating features that facilitate distinguishing anomalies from normal metric segments.
Experiments on three large-scale datasets demonstrate the effectiveness of \tool in faster-than-real-time anomaly anticipation, and cases also show that \tool can make accurate forecasts even under anomalous contexts.
Lastly, we release our code, hoping to provide a useful tool for practitioners and lay the foundation for future research in faster-than-real-time cloud operation.

\section*{Acknowledgement}
The work described in this paper was supported by the Key-Area Research and Development Program of Guangdong Province(No. 2020B010165002) and the Key Program of Fundamental Research from Shenzhen Science and Technology Innovation Commission (No. JCYJ20200109113403826).
It was also supported by the Research Grants Council of the Hong Kong Special Administrative Region, China (No. CUHK 14206921 of the General Research Fund) and the National Natural Science Foundation of China (No. 62202511).

\bibliographystyle{IEEEtran}
\bibliography{main}

% \appendix
% \input{sections/appendix.tex}
\end{document}